\documentclass[lettersize,10pt,fleqn]{article}
\usepackage{graphicx}
\usepackage{subfigure}
\usepackage{morefloats}
\usepackage{color}
\usepackage{setspace}
\usepackage{floatrow}
\usepackage{bm}
\usepackage{color}
\usepackage{float}
\usepackage{amssymb}
\usepackage{geometry}%
\usepackage{amsmath}
\usepackage{CJK} 
\usepackage[colorlinks,linkcolor=blue,citecolor=blue,bookmarks,pdfstartview=FitH]{hyperref}
\usepackage{amsthm,amsmath,amssymb}
\usepackage{mathrsfs}
\usepackage{authblk}
\usepackage{cancel}  
\usepackage{multirow}
\usepackage[bottom]{footmisc}
\usepackage{comment}
\begin{document}
	\begin{sloppypar}
		
		\title{\textbf{Minute-Scale High-Fidelity Gyrokinetic Simulations with Portability from Laptop to Supercomputer}}
		
		\author[1,3]{Jian Bao}
		\author[1]{Huasheng Xie\thanks{huashengxie@gmail.com}}
		\author[1]{Ming Yang}
		\author[2]{Zhixin Lu}
		\author[3]{Haotian Chen}
		\author[3]{Zhihong Lin\thanks{zhihongl@uci.edu}}
		\author[1]{Feng Zhang}
	
   	    \affil[1]{\small Beijing VeloAlpha Technology Co., Ltd., Beijing, China}
   	    \affil[2]{\small Max Planck Institut für Plasmaphysik, 85748 Garching, Germany}
   	    \affil[3]{\small Future Energy Institute, Shanghai Jiao Tong University, Shanghai 200240, China}
	
		\maketitle

	\begin{abstract}
	Global gyrokinetic particle simulations remain computationally expensive, as they demand both adequate marker statistics and three-dimensional field solvers. In this work, we present a hybrid spectral method within the particle-in-Fourier (PIF) framework and implement it in the electrostatic model of GTC. Charge scatter and field gather are performed between particles and fields on a two-dimensional poloidal mesh, while the corresponding Poisson solver is discretized using radial finite differences and poloidal $m$-harmonics. Truncated spectral transforms are employed to connect multiple representations for fields, avoiding costly particle-grid operations for each individual $m$-harmonic within the particle loop. Benchmarks against conventional particle-in-cell (PIC) simulations successfully reproduce single-$n$ ion temperature gradient (ITG) mode structures and dispersion relations, as well as multi-$n$ nonlinear ITG transport and its regulation by zonal flows. Compared to conventional PIC, the proposed method reduces the effective problem size by more than a factor of 48 and achieves a speedup of over two orders of magnitude for single-$n$ cases. A 2000-step single-$n$ simulation with approximately 2 million markers completes in 78.2 seconds on a laptop GPU, while multi-$n$ turbulence simulation also completes within minutes. Furthermore, the elimination of toroidal particle-shift communication yields promising preliminary scaling performance on multiple NVIDIA A100 GPUs. The numerical scheme is broadly applicable for accelerating particle simulations on platforms ranging from laptops to supercomputers.
	\end{abstract}

\section{Introduction}

Gyrokinetic simulation has been widely used as a first-principle approach for turbulent transport in magnetized fusion plasmas, with predictive capability across various plasma regimes \cite{Lee1983}. The two main numerical approaches are particle-in-cell (PIC) and continuum methods. The PIC method is used by GTC \cite{Lin98}, XGC1 \cite{Ku2009}, GEM \cite{Chen2003}, EUTERPE \cite{Kleiber2024} and ORB5 \cite{Lanti2020}, and the continuum method is used by GYRO \cite{Candy2003}, GENE \cite{Gorler2011}, and NLT \cite{Ye2016}. Specifically, PIC method represents the distribution function using Lagrangian markers that move in three-dimensional real space, whereas continuum method discretizes the distribution function in five-dimensional phase space and generally requires greater computational and memory resources.

Statistical noise exists in particle simulations and a large marker population is often required to decrease the noise influence. The $\delta f$-method effectively avoids the noise problem by evolving only the perturbed distribution function \cite{Dimits1993,Parker1993}, rather than the total distribution function. This eliminates the statistical noise that would otherwise arise from the large equilibrium component, allowing for accurate measurements of small perturbed quantities with much fewer markers compared to full-$f$ approach. GTC combines PIC method with comprehensive gyrokinetic physics models in toroidal plasmas \cite{Lee1983,Holod2009,Wei2026}, its GPU implementation has achieved substantial acceleration on top supercomputers such as Summit \cite{Zhang2018}. Meanwhile, The spectral simulation method offers another route to achieve higher efficiency. The particle-in-Fourier (PIF) approach retains the parallel computing advantage of PIC method while reducing the dimensionality of the field representation. It has been verified in simplified geometries \cite{Mitchell2019,Ameres_phd}, extended to multiple-mode calculations \cite{Muralikrishnan2026}, and applied in TRIMEG gyrokinetic code with toroidal \cite{Lu2019} and poloidal \cite{Lu2023} spectral representations, and TRIMEG also develops the field solver in a hybrid finite-element/spectral space \cite{Lu2021}.

However, production-level gyrokinetic simulations still rely heavily on supercomputing resources, which limits the feasibility of extensive parameter scans and the investigation of cross-scale frontier problems. For instance, the GTC code has recently extended its physics models and geometrical capabilities to address radio-frequency waves \cite{Bao14,Bao16,Bao16b,Chen19}, turbulent transport in field-reversed configurations \cite{Bao2019,Wang2021,Sun2020} and in stellarators \cite{Chenht25}. Concurrently, GTC has been employed to study key physics issues such as energetic-particle-driven Alfvén eigenmodes \cite{Taimourzadeh19,Chenl25} and fishbone instabilities \cite{Brochard24}, kinetic ballooning mode (KBM) turbulence \cite{Dong17,Dong19,Xie16} and edge pedestal transport \cite{Xie17}. These diverse applications require both high numerical accuracy and high grid resolution with a large number of markers. Therefore, an efficient, high-fidelity simulation tool is highly desirable to enable the extension of such studies to broader parameter regimes and multiple physical scales.

In this work, we propose a hybrid spectral method within the particle-in-Fourier (PIF) framework and implement it for the electrostatic model in GTC. In this approach, charge scatter and field gather are performed between particle and toroidal Fourier coefficients of fields on a two-dimensional poloidal mesh. The Poisson solver, meanwhile, retains only the physically relevant coupled $m$-harmonics within a sparse matrix constructed by radial finite difference and poloidal $m$-harmonic decomposition. Truncated spectral transforms are employed to connect different field representations and reconstruct the fields on the poloidal plane, avoiding tedious particle-grid operations for each individual $m$-harmonic. The resulting hybrid spectral PIF method reduces the effective problem size by more than a factor of 48 and achieves a speedup of over two orders of magnitude for single-$n$ cases, without sacrificing any physical fidelity. Furthermore, it eliminates the toroidal particle-shift communication typically required in conventional PIC schemes \cite{Either2005}, making it particularly well suited for multi-GPU and multi-MPI rank simulations, as well as future cross-scale studies of microturbulence, energetic-particle-driven instabilities, and macroscopic MHD modes.

The reduced computational cost brings initial-value particle simulations closer to the efficiency of eigenvalue solvers \cite{Xie17comparison,Bao23}, and may facilitate the development of gyrokinetic stability analysis tools. The remainder of this paper is organized as follows. Section 2 presents the physical model and the hybrid spectral method. Section 3 provides numerical verification and performance results. Section 4 concludes the work with a summary and discussion.

\section{Physical model and numerical method}\label{section2}

\subsection{Electrostatic gyrokinetic model}

We adopt the electrostatic gyrokinetic-Poisson system in this work \cite{Lin98}. The gyrocenter dynamics are described by the gyrokinetic equation, with the gyrocenter position $\bm R$, magnetic moment $\mu$, and parallel velocity $v_{||}$ as independent variables in five-dimensional phase space:
\begin{flalign}\label{GK_vlasov}
\left(\frac{\partial}{\partial t} + \dot{\bm R} \cdot\nabla +\dot{v_{||}}\frac{\partial}{\partial v_{||}}\right)f_\alpha\left({\bm R}, v_{||}, \mu, t\right)=0,
\end{flalign}
\begin{flalign}\label{R}
\dot{\bm R}=v_{||}{\bm b} + {\bm {v_{E}}} + {\bm {{v_d}}},
\end{flalign}
\begin{flalign}\label{vpara}
\dot{v_{||}}=-\frac{1}{m_\alpha}\frac{\bm B^*}{B_{||}^*}\cdot\left(Z_\alpha\nabla\left\langle\delta\phi\right\rangle+\mu\nabla B\right),
\end{flalign}
Here, $Z_\alpha$, $m_\alpha$, and $f_\alpha$ are the charge, mass, and distribution function of species $\alpha$, respectively. The equilibrium magnetic field is $\bm B$, with $\bm b={\bm B}/B$, $\bm B^*=\bm B+\left(Bv_{||}/\Omega_{c\alpha}\right)\nabla\times{\bm b}$, and $B_{||}^*={\bm b}\cdot{\bm B^*}$. The electrostatic potential perturbation is $\delta\phi$. The operator $\left\langle\cdots\right\rangle=\left(1/2\pi\right)\int {\bm {dx}}d\xi\left(\cdots\right)\delta\left({\bm R}+{\bm\rho_\alpha}-{\bm x}\right)$ denotes the gyro-phase average, where $\xi$ is the gyro-phase angle, $\bm x$ is the particle position, $\bm\rho_\alpha=\bm b\times\bm v_\perp/\Omega_{c\alpha}$ is the gyroradius vector, and $\Omega_{c\alpha}$ is the cyclotron frequency. The $\bm E\times\bm B$ velocity $\bm v_E$ and magnetic drift velocity $\bm v_d$ are
\begin{flalign*}
{\bm{ v_E}}=\frac{c{\bm b}\times \nabla\left\langle\delta \phi\right\rangle}{B_{||}^*},
\end{flalign*}
and 
\begin{flalign*}
{\bm {v_d}}=\frac{cm_\alpha v_{||}^2}{Z_\alpha B_{||}^*}{\bm b}\times\left({\bm b}\cdot\nabla{\bm b}\right) +
	\frac{c\mu}{Z_\alpha B_{||}^*}{\bm b}\times \nabla B.
\end{flalign*}

To reduce statistical noise, we employ the perturbative $\delta f$-method \cite{Dimits1993,Parker1993}. The distribution function is decomposed into equilibrium and perturbed components, $f_\alpha=f_{\alpha0}\left(\bm R,\mu,v_{||}\right)+\delta f_\alpha\left(\bm R,\mu,v_{||},t\right)$, where the equilibrium distribution $f_{\alpha0}$ satisfies
\begin{flalign}\label{eq_vlasov}
L_0f_{\alpha0}=0,
\end{flalign}
where $L_0={\partial}/{\partial t}+\left(v_{||}{\bm b}+\bm v_d\right)\cdot\nabla-\left({\mu}/{m_\alpha}\right)\bm B^*\cdot\nabla B/B_{||}^*\left(\partial/\partial v_{||}\right)$. We use the local Maxwellian $f_{\alpha0}=n_{\alpha0}\left(m_\alpha/2\pi T_{\alpha0}\right)^{3/2}\exp\left[-\left(m_\alpha v_{||}^2+2\mu B\right)/2T_{\alpha0}\right]$, where $n_{\alpha0}\left(\psi\right)$ and $T_{\alpha0}\left(\psi\right)$ are flux functions. Within the gyrokinetic ordering adopted here, $f_{\alpha0}$ is assumed to satisfy Eq. \eqref{eq_vlasov} to the retained order. Subtracting Eq. \eqref{eq_vlasov} from Eq. \eqref{GK_vlasov} then gives the evolution equation for $\delta f_\alpha$:
\begin{flalign}\label{perturb_vlasov}
L\delta f_\alpha=-\delta Lf_{\alpha0},
\end{flalign}
where $\delta L={\bm v_E}\cdot\nabla-\left(Z_\alpha/m_\alpha/B_{||}^*\right){\bm B^*\cdot\nabla\left\langle\delta\phi\right\rangle}\left(\partial/\partial v_{||}\right)$ and $L=L_0+\delta L$. Defining the marker weight as $w_\alpha=\delta f_\alpha/f_\alpha$, Eq. \eqref{perturb_vlasov} gives
\begin{flalign}\label{ion_weight}
	\begin{split}
		\frac{dw_\alpha}{dt}=\left(1-w_\alpha\right)\left[-{\bm {v_E}}\cdot\frac{\nabla f_{\alpha0}}{f_{\alpha0}}\bigg\lvert_{v_\perp}
		-\frac{Z_\alpha v_{||}{\bm b}\cdot\nabla\left\langle\delta\phi\right\rangle}{T_{\alpha0}}-\frac{Z_\alpha}{T_{\alpha0}}\left(\frac{\mu\bm b\times\nabla B}{m_\alpha\Omega_{c\alpha}}+\frac{v_{||}^2}{\Omega_{c\alpha}}{\bm b}\times\left({\bm b}\cdot\nabla{\bm b}\right)\right)\cdot\nabla\left\langle\delta\phi\right\rangle\right],
	\end{split}
\end{flalign}
In deriving Eq. \eqref{ion_weight} from Eq. \eqref{perturb_vlasov}, we use the chain rule $\nabla f_{\alpha0}\lvert_{v_\perp}=\nabla f_{\alpha0}\lvert_\mu+\mu f_{\alpha0}\nabla B/T_{\alpha0}$. The perturbed thermal-ion density is then given by
\begin{flalign}\label{ni_charge}
\left\langle\delta n_i\left(\bm x,t\right)\right\rangle=\int\delta f_i\left(\bm R, \mu, v_{||},t\right) \bm{dv}{\bm {dR}}d\xi\delta\left(\bm R+\bm{\rho_i}-\bm{x}\right)/\left(2\pi\right).
\end{flalign}

The gyrokinetic Poisson equation is \cite{Lee1983}
\begin{flalign}\label{poisson}
\frac{Z_i^2n_{i0}}{T_{i0}}\left(\delta\phi-\widetilde{\delta\phi}\right) + \frac{e^2n_{e0}}{T_{e0}}\delta\phi=Z_i\left\langle\delta n_i\right\rangle,
\end{flalign}
where $n_{i0}$ and $T_{i0}$ are the equilibrium ion density and temperature. The quantity $\widetilde{\delta\phi}\left(\bm x,t\right)=\frac{1}{n_{i0}}\int f_{i0}\left(\bm R,\mu,v_{||}\right)\left\langle\delta\phi\right\rangle{\bm{dv}}\bm{dR}d\xi\,\delta\left(\bm R+\bm\rho_i-\bm x\right)/\left(2\pi\right)$ is the double gyro-phase average of the ion electrostatic potential, and the velocity-space measure is $\int{\bm{dv}}=\frac{2\pi B}{m_i}\int dv_{||}d\mu$. In the following parts of this paper, we omit the gyro-average bracket $\langle\cdots\rangle$ for notation symplicity.

\subsection{Hybrid spectral PIF method}
\begin{figure}[H]
	\center
	\includegraphics[width=1\textwidth]{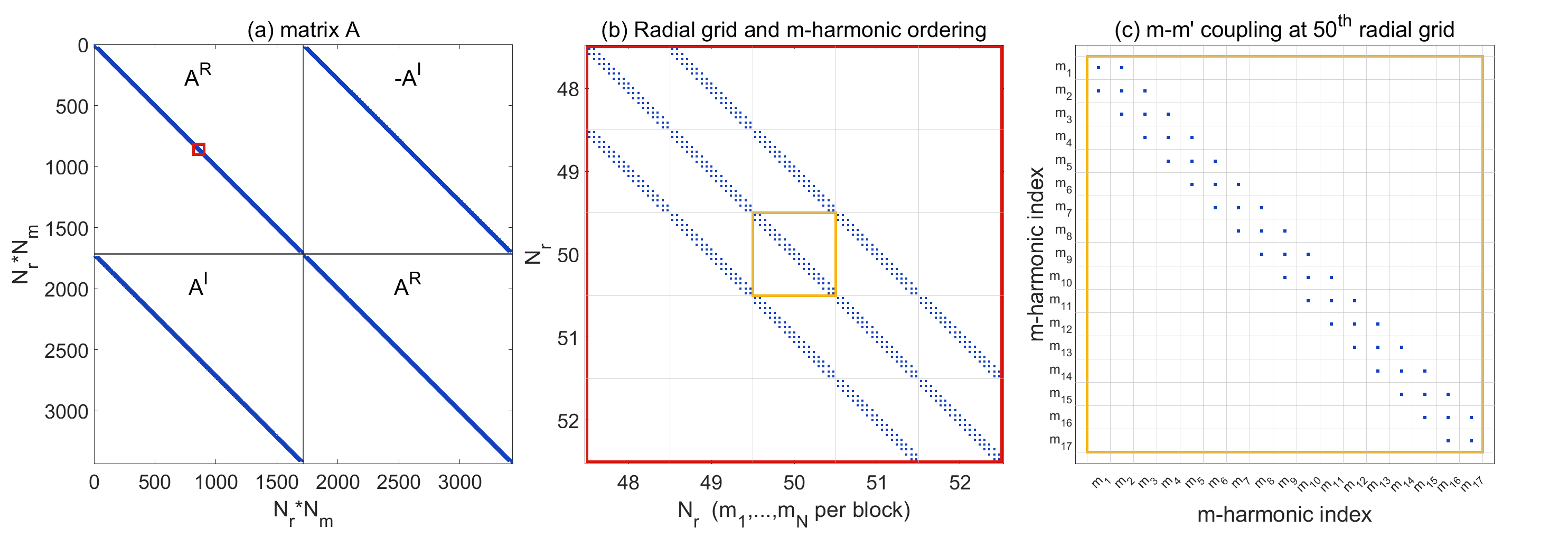}
	\caption{Sparse-matrix structure of the spectral Poisson solver. (a) Real block representation of Eq. \eqref{matA}, where the blue dots denote nonzero elements. (b) Ordering of the radial-grid and $m$-harmonic indices, illustrating the three-point radial finite-difference coupling. (c) Coupling among $m$ harmonics within the highlighted radial block, the three diagonals correspond to coupling between $m$ and $m\pm1$.}
	\label{matrix}	
\end{figure}

GTC evolves particles and fields in three-dimensional real space using Boozer coordinates $\left(\psi,\theta,\zeta\right)$, where $\psi$ is the poloidal magnetic flux and $\theta$ and $\zeta$ are the Boozer poloidal and toroidal angles, respectively. Here we extend the mixed PIC-PIF method \cite{Lu2019,Lu2023}, originally developed with a finite-element field solver, to the finite-difference scheme in GTC. Different spectral representations are used for charge scatter, field gather, and the Poisson solver so that each stage of the particle-field cycle can be treated efficiently. With the ansatz $\exp\left(-im\theta+in\zeta\right)$, the thermal-ion density perturbation $\delta n_i$ is written as
\begin{flalign}\label{n_expand}
	\begin{split}
		\delta n\left(\psi,\theta,\zeta,t\right) 
		= &\sum_m\delta n_{m,n}\left(\psi,t\right)\exp\left(-im\theta + i n \zeta\right)
		+ \sum_m\delta n_{m,n}^*\left(\psi,t\right)\exp\left(im\theta - i n \zeta\right)\\
		=&2\sum_{m}\delta n_{m,n}^R\left(\psi,t\right)\cos\left(-m\theta+n\zeta\right)
		-2\sum_{m}\delta n_{m,n}^I\left(\psi,t\right)\sin\left(-m\theta+n\zeta\right)
	\end{split}
\end{flalign}
where $\delta n_{m,n} = \delta n_{m,n}^R + i\delta n_{m,n}^I$ and $\delta n_{m,n}^* = \delta n_{m,n}^R - i\delta n_{m,n}^I$. The electrostatic potential $\delta\phi$ can be expressed as 
\begin{flalign}\label{phi_expand}
	\begin{split}
		\delta\phi\left(\psi,\theta,\zeta,t\right) 
		= &\sum_m\delta\phi_{m,n}\left(\psi,t\right)\exp\left(-im\theta + i n \zeta\right)
		+ \sum_m\delta\phi_{m,n}^*\left(\psi,t\right)\exp\left(im\theta - i n \zeta\right)\\
		=&2\sum_{m}\delta\phi_{m,n}^R\left(\psi,t\right)\cos\left(-m\theta+n\zeta\right)
		-2\sum_{m}\delta\phi_{m,n}^I\left(\psi,t\right)\sin\left(-m\theta+n\zeta\right)
	\end{split}
\end{flalign}
where $\delta\phi_{m,n}=\delta\phi_{m,n}^R+i\delta\phi_{m,n}^I$ and $\delta\phi_{m,n}^*=\delta\phi_{m,n}^R-i\delta\phi_{m,n}^I$. Equation \eqref{poisson} can then be written in matrix form as
\begin{equation}\label{matA}
   \left[
	\begin{array}{cc}
		\mathbb{A}^R& -\mathbb{A}^I\\
		\mathbb{A}^I& \mathbb{A}^R\\
	\end{array}
	\right ]\left[
	\begin{array}{c}
		\boldsymbol{\delta\phi}^R\\
		\boldsymbol{\delta\phi}^I\\
	\end{array}
	\right]=\left[
	\begin{array}{c}
		\boldsymbol{\delta n}^R\\
		\boldsymbol{\delta n}^I\\
	\end{array}
	\right],
\end{equation}
Here, $\mathbb{A}=\mathbb{A}^R+i\mathbb{A}^I$ is the complex Poisson matrix. The vectors $\boldsymbol{\delta\phi}^R=[\cdots,\delta\phi_{m-1,n}^R(i),\delta\phi_{m,n}^R(i),\delta\phi_{m+1,n}^R(i),\cdots]^T$ and $\boldsymbol{\delta\phi}^I=[\cdots,\delta\phi_{m-1,n}^I(i),\delta\phi_{m,n}^I(i),\delta\phi_{m+1,n}^I(i),\cdots]^T$ contain the real and imaginary potential coefficients; $\boldsymbol{\delta n}^R$ and $\boldsymbol{\delta n}^I$ are ordered in the same way. The index $i$ denotes the radial grid point. Figure \ref{matrix} shows the block structure of Eq. \eqref{matA}. Unlike the traditional finite-difference approach defined on a two-dimensional poloidal grid \cite{Xiao15}, the radial‑$m$ decomposition adopted here reduces the matrix dimension by more than two orders of magnitude while improving accuracy. This is achieved by applying a spectral discretization in the poloidal direction and retaining only the physically dominant $m$-$m$ couplings, thereby effectively reducing the dimensionality of the problem. 

The Fourier representation is retained only toroidally for both field gather and charge scatter, which are carried out between particles and 2D poloidal plane grids. This choice eliminates the cost of carrying out particle-grid operations for each individual $m$-harmonic within the particle loop. The electrostatic potential is therefore expressed as
\begin{flalign}\label{phi_n}
	\begin{split}
		\delta\phi\left(\psi,\theta,\zeta,t\right) 
		=&\delta\phi_n\left(\psi,\theta,t\right)e^{in\zeta} +     \delta\phi_n^*\left(\psi,\theta,t\right)e^{-in\zeta}\\
		=&2\left[\delta\phi_n^R\left(\psi,\theta,t\right)\cos\left(n\zeta\right)
		-\delta\phi_n^I\left(\psi,\theta,t\right)\sin\left(n\zeta\right)\right]
	\end{split}
\end{flalign}
where $\delta\phi_n=\delta\phi_n^R+i\delta\phi_n^I$ and $\delta\phi_n^*=\delta\phi_n^R-i\delta\phi_n^I$. The field gradients required by the particle pusher are evaluated in real space from $\delta\phi_n^R$ and $\delta\phi_n^I$ as
\begin{flalign}\label{dphidp}
	\begin{split}
		\frac{\partial\delta\phi}{\partial\psi}
		=2\left[\frac{\partial\delta\phi_n^R}{\partial\psi}\cos\left(n\zeta\right)
		-\frac{\partial\delta\phi_n^I}{\partial\psi}\sin\left(n\zeta\right)\right]
	\end{split}
\end{flalign}
\begin{flalign}\label{dphidt}
	\begin{split}
		\frac{\partial\delta\phi}{\partial\theta}
		=2\left[\frac{\partial\delta\phi_n^R}{\partial\theta}\cos\left(n\zeta\right)
		-\frac{\partial\delta\phi_n^I}{\partial\theta}\sin\left(n\zeta\right)\right]
	\end{split}
\end{flalign}
\begin{flalign}\label{dphidz}
	\begin{split}
		\frac{\partial\delta\phi}{\partial\zeta}
		=2\left[-n\delta\phi_n^R\sin\left(n\zeta\right)
		-n\delta\phi_n^I\cos\left(n\zeta\right)\right]
	\end{split}
\end{flalign}
The solution $\delta\phi_{m,n}^{R,(I)}$ of Eq. \eqref{matA} is transformed into $\delta\phi_n^{R,(I)}$ and its gradients in Eqs. \eqref{dphidp}--\eqref{dphidz}. Only the dominant $m$ harmonics are retained in the spectral sums used to reconstruct the fields on each poloidal plane:
\begin{flalign}\label{phi_n_R}
	\begin{split}
		 \delta\phi_n^R = \sum_{m}\left[\delta\phi_{m,n}^R\cos\left(m\theta\right) + \delta\phi_{m,n}^I\sin\left(m\theta\right)\right]
	\end{split},
\end{flalign}
\begin{flalign}\label{phi_n_I}
	\begin{split}
		\delta\phi_n^I = \sum_{m}\left[-\delta\phi_{m,n}^R\sin\left(m\theta\right) + \delta\phi_{m,n}^I\cos\left(m\theta\right)\right]
	\end{split}
\end{flalign}
\begin{flalign}\label{dphidp_n_R}
	\begin{split}
		\frac{\partial\delta\phi_n^R}{\partial\psi} = \sum_{m}\left[\frac{\partial\delta\phi_{m,n}^R}{\partial\psi}\cos\left(m\theta\right) + \frac{\partial\delta\phi_{m,n}^I}{\partial\psi}\sin\left(m\theta\right)\right]
	\end{split},
\end{flalign}
\begin{flalign}\label{dphidp_n_I}
	\begin{split}
		\frac{\partial \delta\phi_n^I}{\partial\psi}= \sum_{m}\left[-\frac{\partial \delta\phi_{m,n}^R}{\partial\psi}\sin\left(m\theta\right) + \frac{\partial \delta\phi_{m,n}^I}{\partial\psi}\cos\left(m\theta\right)\right]
	\end{split}
\end{flalign}
\begin{flalign}\label{dphidt_n_R}
	\begin{split}
		\frac{\partial\delta\phi_n^R}{\partial\theta}= \sum_{m}m\left[-\delta\phi_{m,n}^R\sin\left(m\theta\right) + \delta\phi_{m,n}^I\cos\left(m\theta\right)\right]
	\end{split},
\end{flalign}
\begin{flalign}\label{dphidt_n_I}
	\begin{split}
		\frac{\partial \delta\phi_n^I}{\partial\theta} = \sum_{m}m\left[-\delta\phi_{m,n}^R\cos\left(m\theta\right) - \delta\phi_{m,n}^I\sin\left(m\theta\right)\right]
	\end{split}
\end{flalign}
In addition to field gathering, charge scattering is also performed on the 2D $\left(\psi,\theta\right)$ poloidal plane grids. The density perturbation is represented spectrally only in the toroidal direction:
\begin{flalign}\label{n_n}
	\begin{split}
		\delta n\left(\psi,\theta,\zeta,t\right) 
		=&\delta n_n\left(\psi,\theta,t\right)e^{in\zeta} +     \delta n_n^*\left(\psi,\theta,t\right)e^{-in\zeta}\\
		=&2\left[\delta n_n^R\left(\psi,\theta,t\right)\cos\left(n\zeta\right)
		-\delta n_n^I\left(\psi,\theta,t\right)\sin\left(n\zeta\right)\right]
	\end{split}
\end{flalign}
Following the PIF formulation in Ref. \cite{Lu2019}, the complex Fourier coefficient $\delta n_n$ is related to the marker weight $w=\delta f/f$. Equation \eqref{ni_charge} then becomes
\begin{flalign}\label{n_n_scatter}
	\begin{split}
		\frac{\delta n_n\left(\psi,\theta,t\right)}{\langle n \rangle}=\frac{V_{tot}}{N\Delta V}\sum_{\left(\psi,\theta\right)\in\Delta S}^{} w\times e^{-in\zeta}
	\end{split}
\end{flalign}
where $\Delta S$ is the projection of $\Delta V$ onto the $\left(\psi,\theta\right)$ plane. The real and imaginary components are
\begin{flalign}\label{n_n_R}
	\begin{split}
		\frac{\delta n_n^R\left(\psi,\theta,t\right)}{\langle n \rangle}=\frac{V_{tot}}{N\Delta V}\sum_{\left(\psi,\theta\right)\in\Delta S}^{} w\times \cos\left(n\zeta\right)
	\end{split}
\end{flalign}
and 
\begin{flalign}\label{n_n_I}
	\begin{split}
		\frac{\delta n_n^I\left(\psi,\theta,t\right)}{\langle n \rangle}= - \frac{V_{tot}}{N\Delta V}\sum_{\left(\psi,\theta\right)\in\Delta S}^{} w\times \sin\left(n\zeta\right)
	\end{split}
\end{flalign}

To construct the source term of the gyrokinetic Poisson equation in Eq. \eqref{matA}, the dominant coefficients $\delta n_{m,n}^{R,(I)}$ are projected efficiently from $\delta n_n^{R,(I)}$ using the truncated spectral representation:
\begin{flalign}\label{n_mn_R}
	\begin{split}
	     \delta n_{m,n}^R\left(\psi,t\right) = \frac{1}{2\pi}\int\left[\delta n_n^R\cos\left(m\theta\right) - \delta n_n^I\sin\left(m\theta\right)\right]d\theta 
	\end{split}
\end{flalign}
and 
\begin{flalign}\label{n_mn_I}
	\begin{split}
		\delta n_{m,n}^I\left(\psi,t\right) = \frac{1}{2\pi}\int\left[\delta n_n^R\sin\left(m\theta\right) + \delta n_n^I\cos\left(m\theta\right)\right]d\theta 
	\end{split}
\end{flalign}

Equations \eqref{GK_vlasov}--\eqref{vpara}, \eqref{ion_weight}, \eqref{matA}, \eqref{dphidp}--\eqref{dphidt_n_I}, and \eqref{n_n_R}--\eqref{n_mn_I} constitute a closed electrostatic gyrokinetic system formulated within the hybrid spectral PIF framework. The associated reduction in computational cost arises from three main sources. First, the toroidal grid dimension and particle-shift communication are eliminated from the three-dimensional particle-grid coupling, which is particularly well suited for multi-GPU and multi-MPI rank particle simulations. Second, the Poisson matrix retains only the physically relevant harmonics $m\sim nq$ and a limited set of couplings $m\pm\Delta m$, thereby preserving sparsity while substantially reducing the matrix size. Third, field gathering and charge scattering are performed using intermediate toroidal Fourier coefficients on the poloidal plane, which avoids costly particle-grid operations for each individual $m$-harmonic, and the use of truncated spectral transforms are much more efficient than FFT with keeping full spectra.

\section{Verification and numerical performance}
We carry out the physics verification for hybrid spectral PIF method using a concentric circular equilibrium based on Cyclone Base Case (CBC) parameters. The geometry is specified by the major radius $R_0=83.5~\mathrm{cm}$, inverse aspect ratio $a/R_0=0.357$, safety factor $q=1.4$, magnetic shear $s=\left(r/q\right)\left(dq/dr\right)=0.78$ at $r=0.5a$, and on-axis magnetic field $B_a=2.01~\mathrm{T}$. The plasma parameters are $T_i=T_e=2223~\mathrm{eV}$, $R_0/L_{T,i}=R_0/L_{T,e}=6.9$, and $R_0/L_{n,e}=2.2$, where $L_{T,e}=|\nabla T_e/T_e|^{-1}$, $L_{T,i}=|\nabla T_i/T_i|^{-1}$, and $L_{n,e}=|\nabla n_e/n_e|^{-1}$.

\begin{figure}[H]
	\center
	\includegraphics[width=1.0\textwidth]{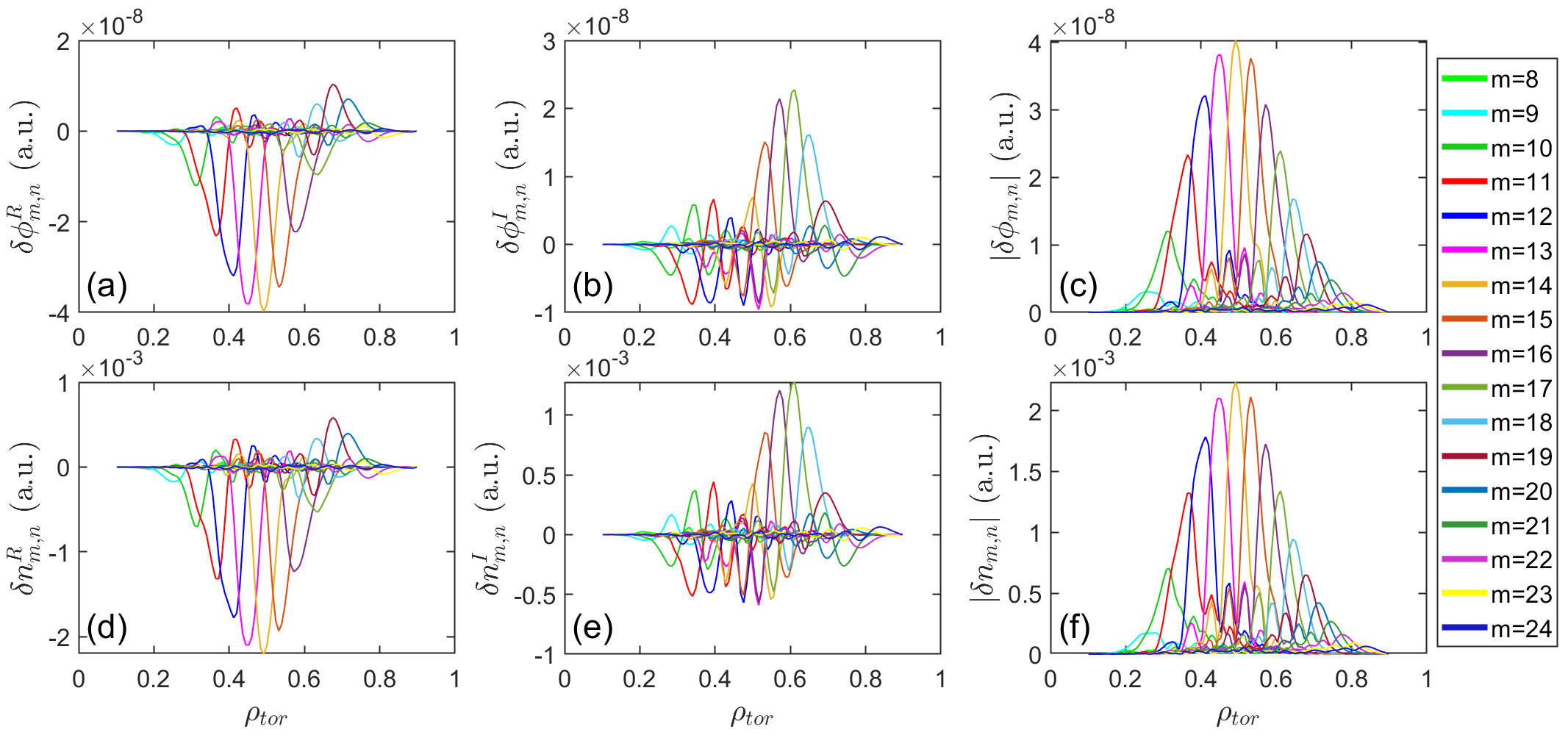}
	\caption{Radial structures of the $m$ harmonics for the $n=10$ ITG mode. (a)--(c) Real, imaginary, and absolute values of the electrostatic-potential coefficients $\delta\phi_{m,n}$ in Eq. \eqref{phi_expand}. (d)--(f) Corresponding density-perturbation coefficients $\delta n_{m,n}$ in Eq. \eqref{n_expand}.}
	\label{snh}	
\end{figure}

\subsection{Single-$n$ simulation}
\begin{figure}[H]
	\center
	\includegraphics[width=1.0\textwidth]{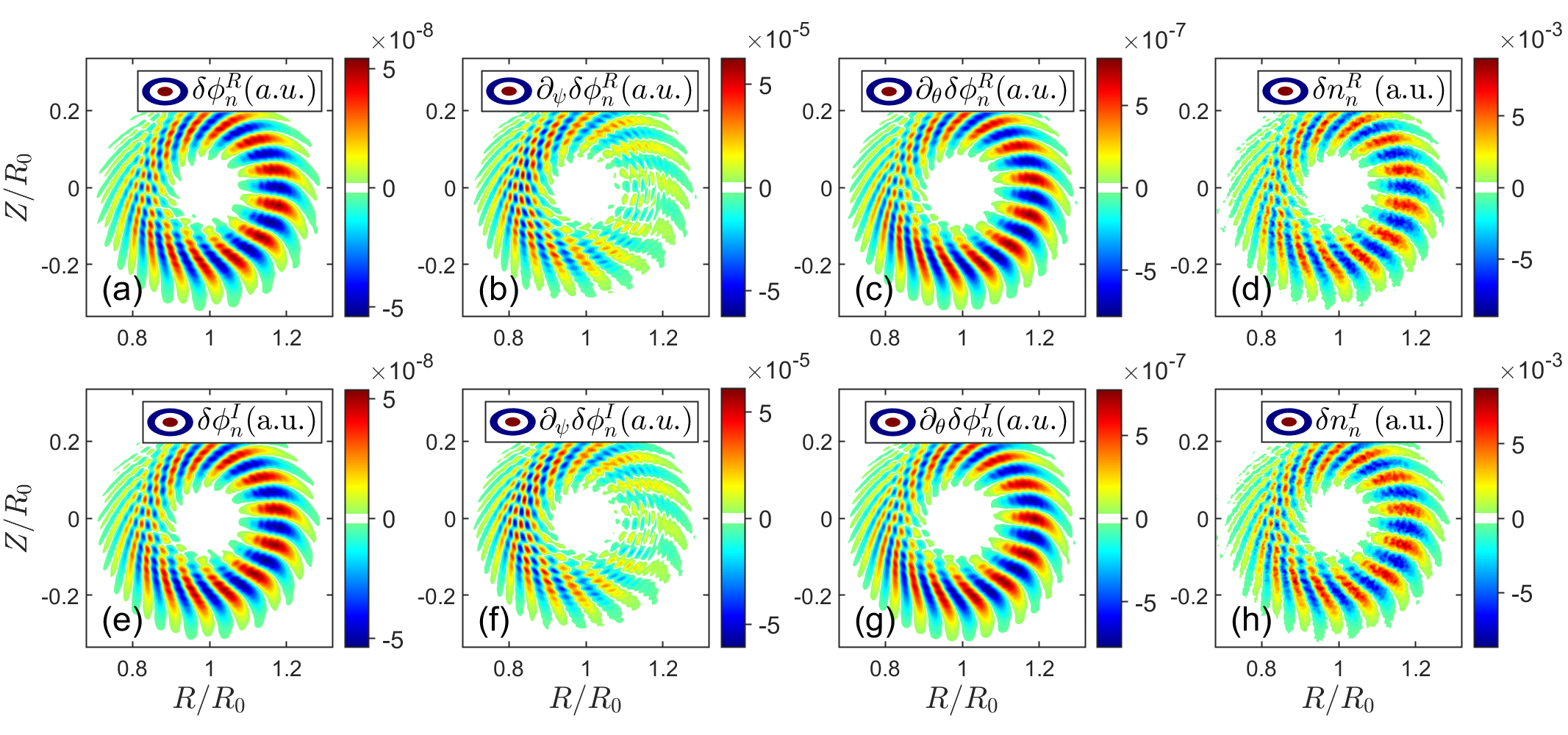}
	\caption{Two-dimensional $n=10$ ITG toroidal Fourier coefficients for particle-grid gather and scatter processes. From left to right, the columns show $\delta\phi_n$, $\partial_\psi\delta\phi_n$, $\partial_\theta\delta\phi_n$, and $\delta n_n$. The upper and lower rows show the real and imaginary components, respectively.}
	\label{snv}	
\end{figure}

\begin{figure}[H]
	\center
	\includegraphics[width=0.75\textwidth]{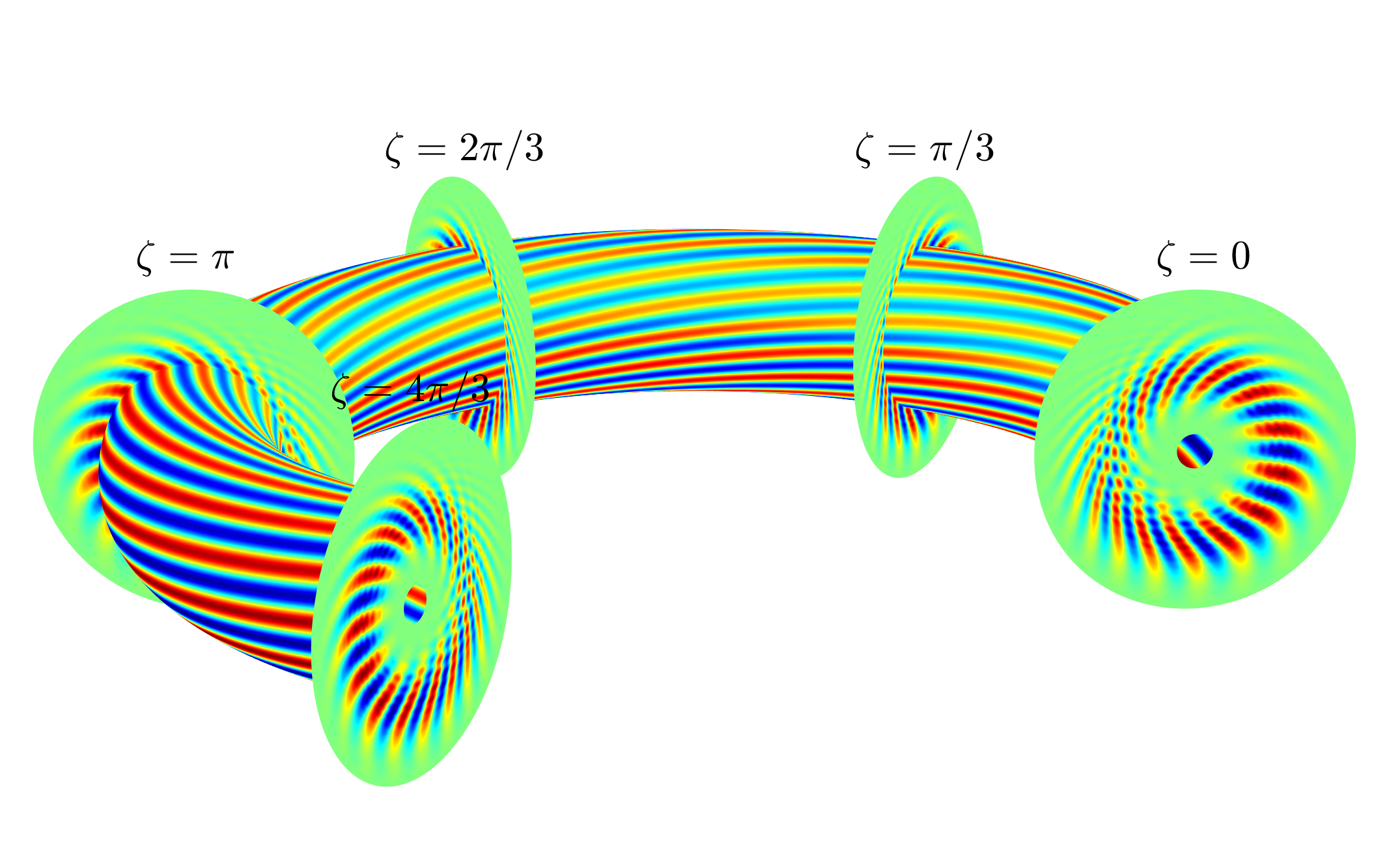}
	\caption{Three-dimensional $n=10$ ITG mode structure of $\delta\phi$ in Cartesian coordinates, reconstructed from $\delta\phi_n$ using Eq. \eqref{phi_n}.}
	\label{sn3d}	
\end{figure}

We first apply the hybrid spectral PIF method to simulate linear $n=10$ ITG mode. Figure \ref{snh} shows the radial structures of the $m$-harmonics of $\delta\phi_{m,n}$ and $\delta n_{m,n}$, which form the solution and source vectors, respectively, of the Poisson solver matrix described by Eq. \eqref{matA}. It is seen that both $|\delta\phi_{m,n}|$ and $|\delta n_{m,n}|$ exhibit typical ballooning structures in figures \ref{snh} (c) and (f) respectively. Charge scatter deposits $\delta n_n^{R,(I)}$ on the 2D $\left(\psi,\theta\right)$ mesh from marker particles as shown by figures \ref{snv}(d) and (h), which are then transformed to $\delta n_{m,n}^{R,(I)}$ in figures \ref{snh} (d) and (e) based on the truncated spectral transforms described by Eqs. \eqref{n_mn_R} and \eqref{n_mn_I}. Conversely, Eqs. \eqref{phi_n_R}--\eqref{dphidt_n_I} reconstruct $\delta\phi_n^{R,(I)}$ in figures \ref{snv} (a) and (e) from $\delta\phi_{m,n}^{R,(I)}$ in figures \ref{snh} (a) and (b), as well as corresponding radial and poloidal derivatives on 2D poloidal plane for field gather process. Then the field gradients in real space can be evaluated by Eqs. \eqref{dphidp}-\eqref{dphidz} for particle pusher. It should be mentioned that after solving Poisson's equation, a simple radial smooth and $m$-harmonic select rule $m\sim nq$ baesd on phsics consideration can effectively suppress nonresonant high-$k$ noise, which is appropriate for most plasma instabilities characterized by $k_{||}\sim0$. The 3D ITG mode structure reconstructed based on Eq. \eqref{phi_n} is shown in figure \ref{sn3d}, which exhibits the typical flute-like mode structure similar to figure 11 of Ref. \cite{Wei2025}. 

\begin{figure}[H]
	\center
	\includegraphics[width=0.8\textwidth]{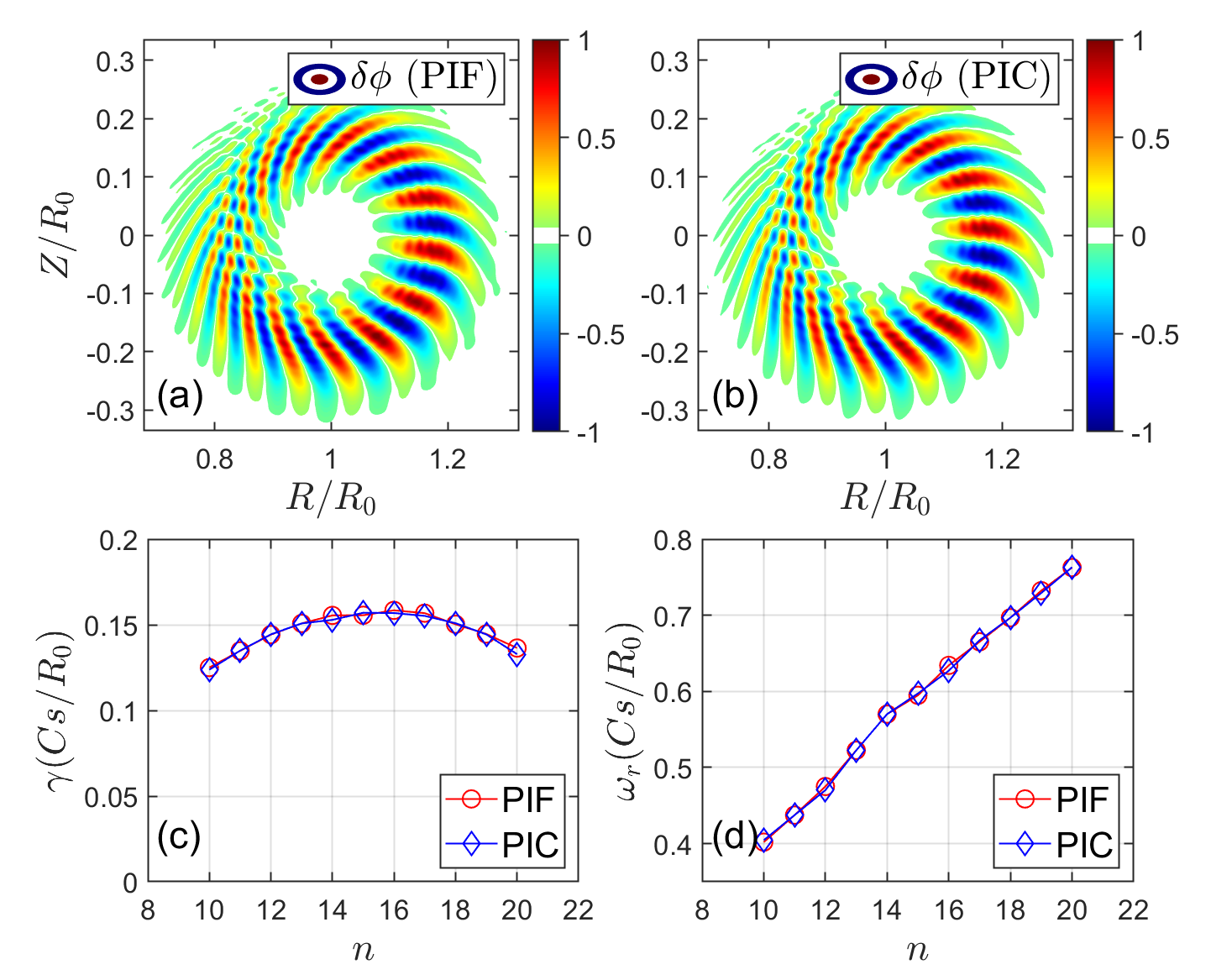}
	\caption{Benchmark of the hybrid spectral PIF method against conventional PIC for linear ITG modes. 2D poloidal mode structures of $\delta\phi$ from (a) hybrid spectral PIF and (b) conventional PIC simulations for $n=10$ case. (c) Growth rate $\gamma$ and (d) real frequency $\omega_r$ for $n=10-20$.}
	\label{snDc}	
\end{figure}

Figure \ref{snDc} compares the hybrid spectral PIF method with conventional PIC for linear ITG dispersion relation and mode structure. The normalized electrostatic-potential structures in figures \ref{snDc}(a) and (b) show excellent agreements on the ballooning structure with nearly identical radial and poloidal patterns. The growth rates and real frequencies also agree with each other for $n=10-20$, including the nonmonotonic variation of the growth rate due to FLR stabilization and the approximately monotonic increase of the real frequency due to increasing diamagnetic drift frequency. The hybrid spectral PIF method therefore faithfully captures the single-$n$ linear ITG physics according to the mode structures and dispersion relations.

The reduction in computational cost mainly comes from replacing the 3D particle-grid system in $(\psi,\theta,\zeta)$ with toroidal Fourier coefficients on a 2D poloidal mesh and radial-$m$ Posson solver. For the single-$n$ case, the hybrid spectral PIF calculation uses approximately 2 million markers, whereas conventional PIC constructs the full toroidal grids and a 50\% larger poloidal grid number for numerical convergence due to real space Poisson solver and smooth, resulting an approximately 97 million markers. The effective problem size is therefore 48 times smaller. Retaining only the physical components of $m$-harmonics and removing nonresonant high-$k$ fluctuations, a clean ITG mode structure can be obtained using 2000 time-step simulation on minute scale on a laptop. Regrading particle convergence, 2 million markers are sufficient to resolve and reconstruct the three-dimensional ITG mode in this case.

\begin{figure}[H]
	\center
	\includegraphics[width=0.9\textwidth]{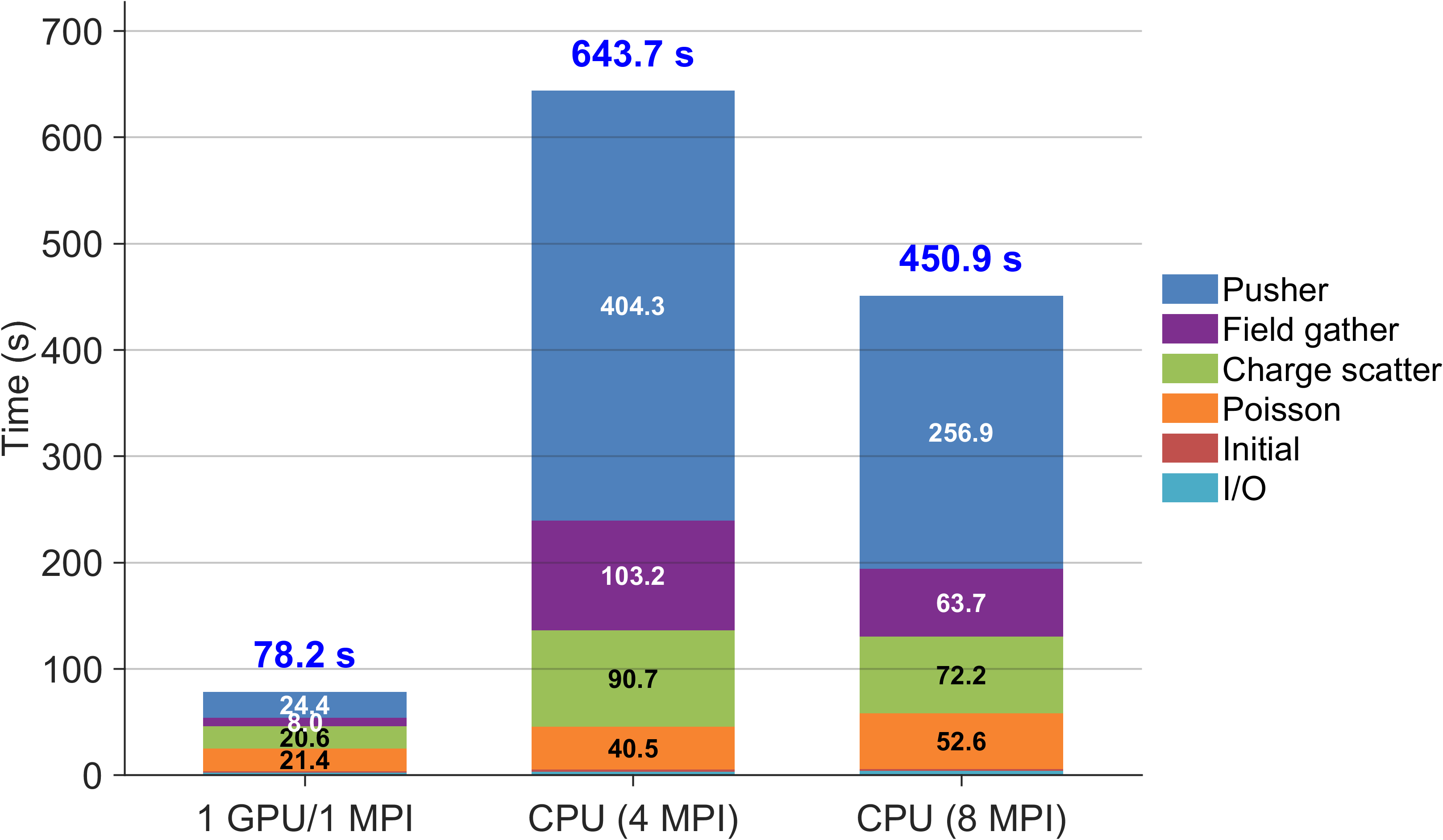}
	\caption{Runtime breakdown for a 2000-step, single-$n$ hybrid spectral PIF simulation with approximately 2 million markers on a laptop. Results are shown for one laptop GPU with one MPI rank and for CPU-only runs with 4 and 8 MPI ranks. The laptop has an Intel Core i9-13900HX processor (24 cores, 32 threads, a 2.2~GHz P-core base frequency, and a 5.4~GHz maximum turbo frequency) and an NVIDIA GeForce RTX 4090 laptop GPU with 9728 CUDA cores and 16~GB of GDDR6 memory.}
	\label{cgtb}	
\end{figure}

Figure \ref{cgtb} gives the runtime breakdown for 2000 time steps of single-$n$ ITG run with 2 million marker on a laptop, which has an Intel Core i9-13900HX processor and an NVIDIA GeForce RTX 4090 laptop GPU. The 1-GPU/1-MPI calculation takes 78.2~s, compared with 643.7 and 450.9~s for CPU-only runs using 4 and 8 MPI ranks, respectively. Considering with the smaller problem size, these timings indicate an overall speedup of more than two orders of magnitude compared to conventional PIC at the same number of markers per cell. Increasing the CPU allocation from 4 to 8 MPI ranks gives further acceleration, and detailed strong-scaling performance and furture optimization will be reported in a separate work. The comprehensive tests for larger marker populations and multiple-$n$ calculations are discussed in Sec. \ref{scaling}.

\subsection{Multiple-$n$ simulations}
\begin{figure}[H]
	\center
	\includegraphics[width=1.0\textwidth]{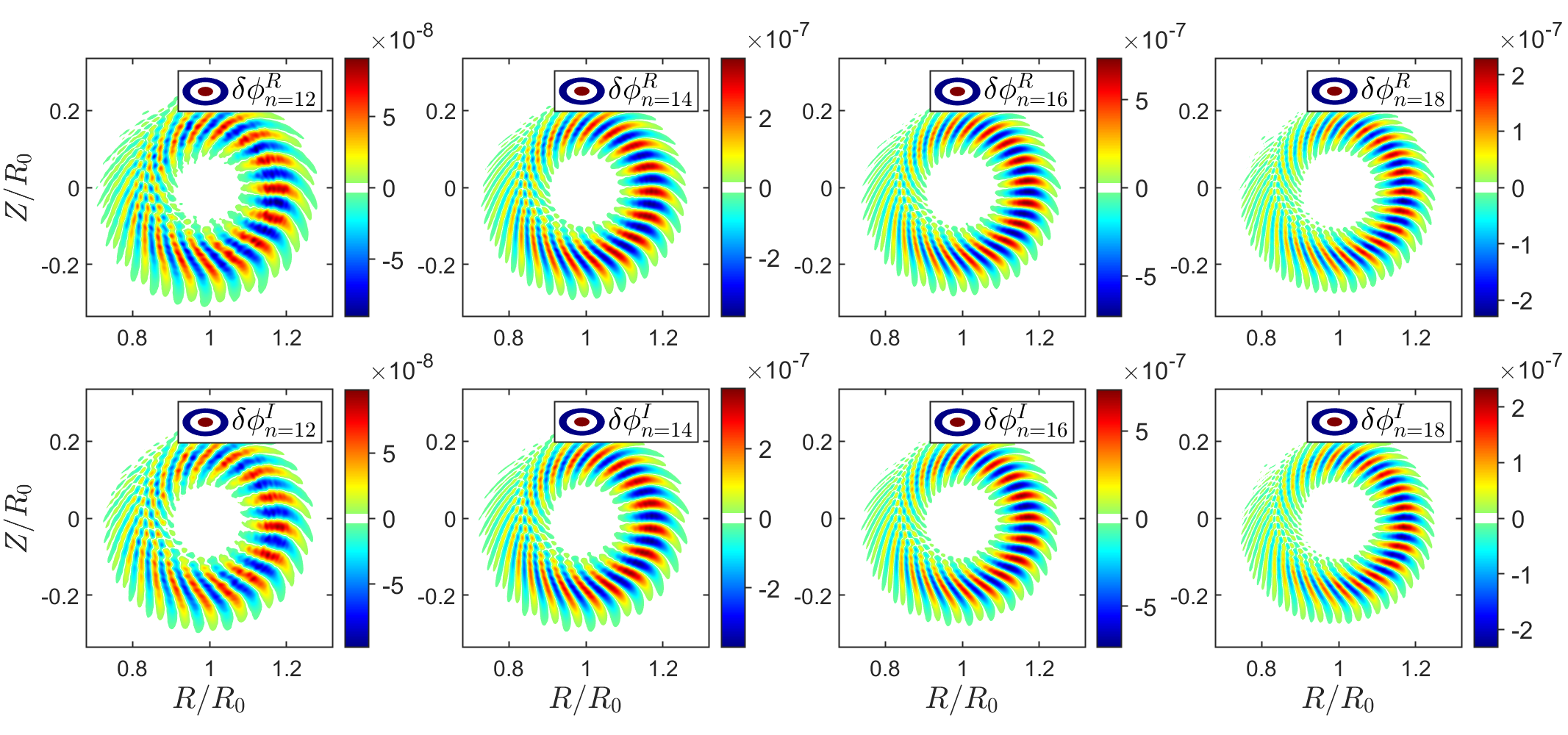}
	\caption{Real (upper row) and imaginary (lower row) components of four dominant toroidal harmonics $\delta\phi_n$ from the linear multiple-$n$ simulation. The columns from left to right correspond to $n$=12, 14, 16 and 18 results.}
	\label{mnh}	
\end{figure}

We next consider a linear multiple-$n$ ITG case retaining the six toroidal harmonics $n=10$, 12, 14, 16, 18, and 20. Figure \ref{mnh} shows the real and imaginary components for dominant $n=12$, 14, 16, and 18. It is seen that each component is clearly resolved and has the outer-midplane localized ballooning structure, in consistency with typical ITG mode characteristics. It should be pointed out that there are some fine-radial scale oscillations on $\delta\phi$ structure in the first column of Fig. \ref{mnh}, which is due to the small growth rate of $n=12$ component with weaker physical signal that can be affected by particle noise, and could be resolved by increasing the marker particle number. The dominant poloidal mode number satisfies $m\simeq nq$, so the poloidal wavelength decreases with increasing $n$ number. The hybrid spectral PIF method retains the mode structure and phase of each toroidal component without storing the fields on 3D real-space mesh.

\begin{figure}[H]
	\center
	\includegraphics[width=0.75\textwidth]{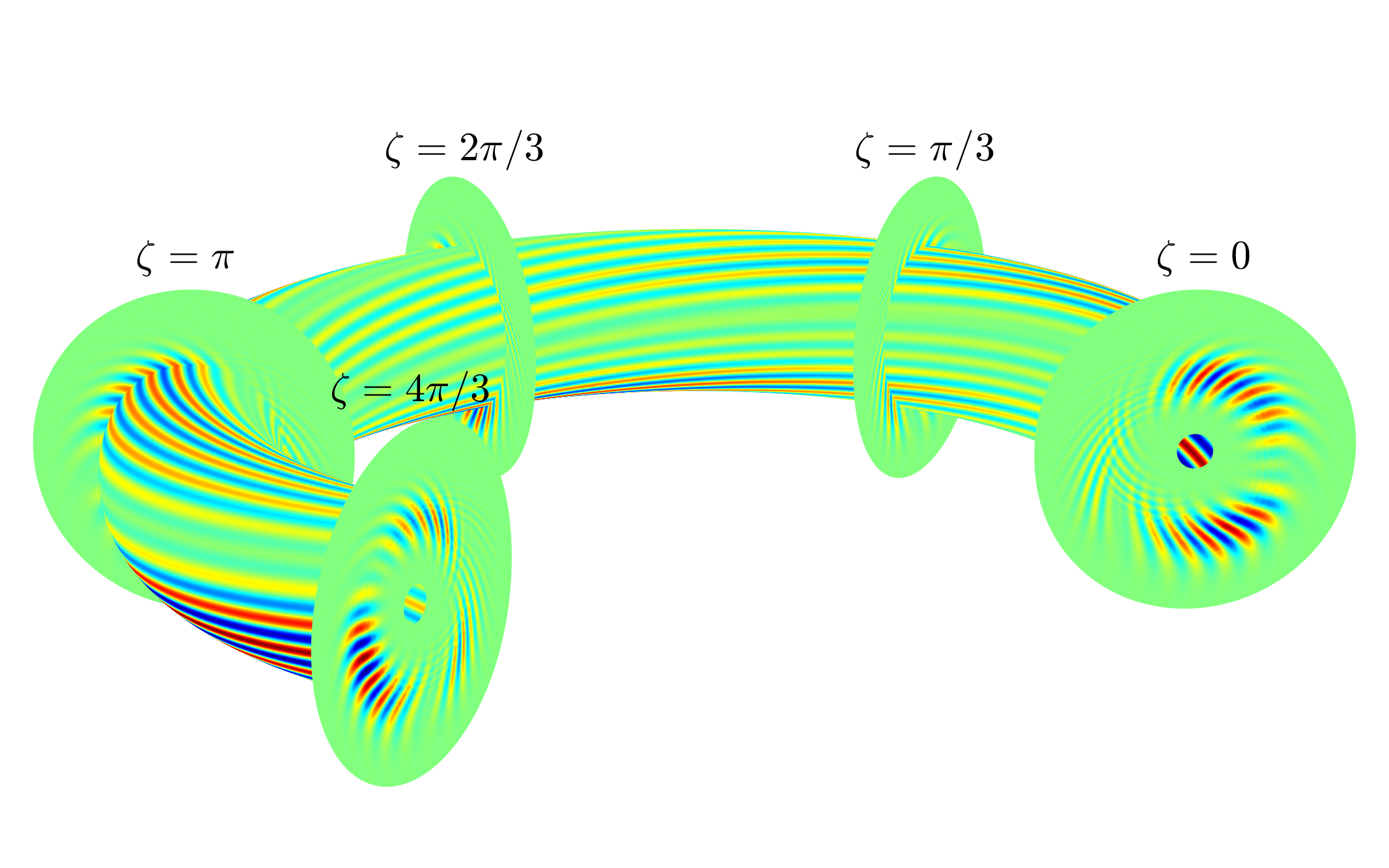}
	\caption{Three-dimensional electrostatic-potential perturbation reconstructed in Cartesian coordinates by summing Eq. \eqref{phi_n} over the retained toroidal harmonics $n=10$, 12, 14, 16, 18, and 20. The labels indicate the toroidal angle $\zeta$.}
	\label{mn3d}	
\end{figure}

\begin{figure}[H]
	\center
	\includegraphics[width=0.9\textwidth]{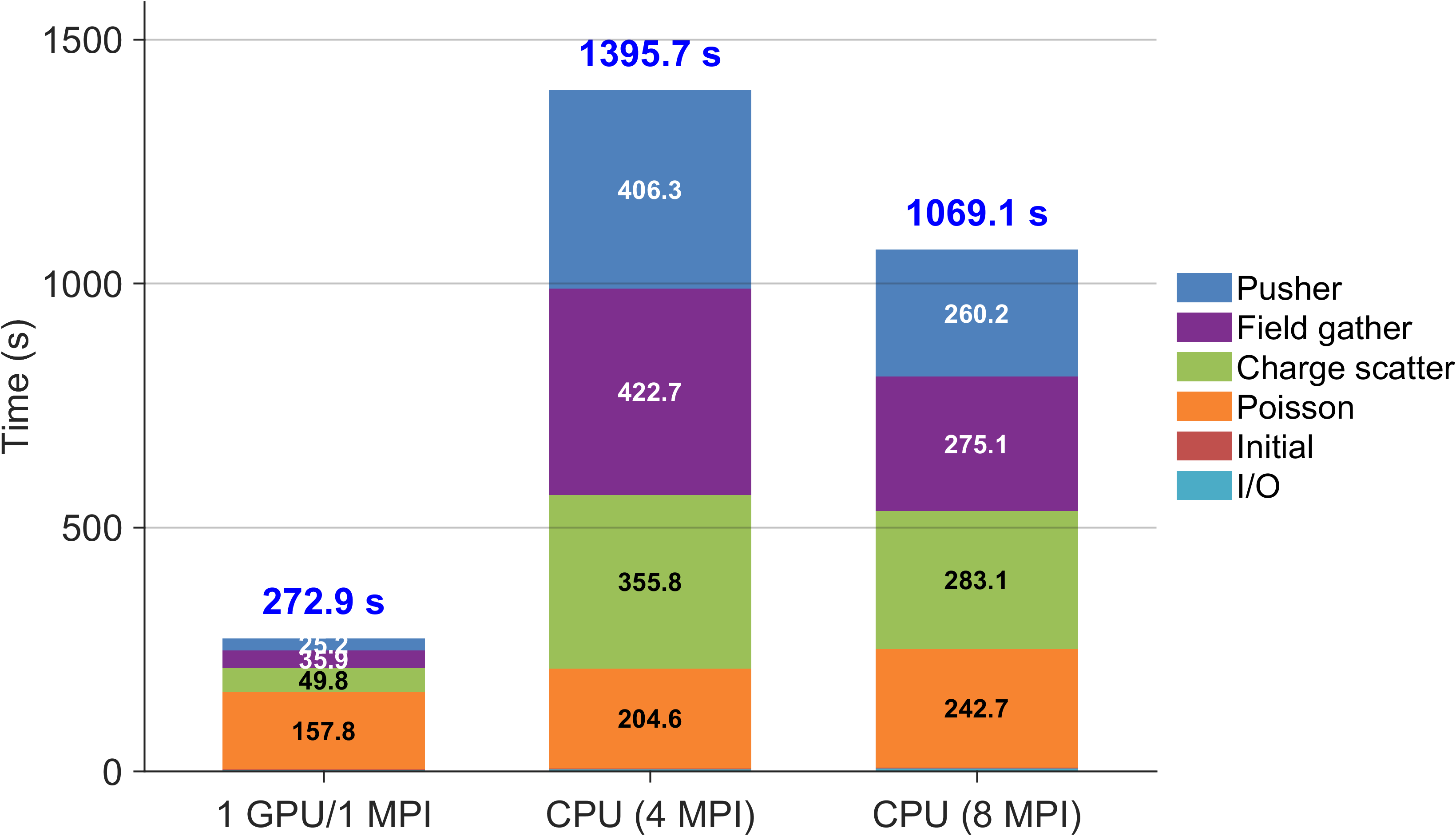}
	\caption{Runtime breakdown of the linear multiple-$n$ simulation with six retained toroidal harmonics for 2000 time steps on a laptop. The bars show the dominant computational components for 1 GPU/1 MPI rank and CPU-only runs with 4 and 8 MPI ranks. The corresponding total runtimes are 272.9, 1395.7 and 1069.1~s, respectively.}
	\label{cgtb_6n}	
\end{figure}

The full 3D $\delta\phi$ perturbation is reconstructed by summing Eq. \eqref{phi_n} over all six toroidal harmonics. In figure \ref{mn3d}, their interference produces a toroidally varying envelope and finer spatial scales than those of an individual mode. The 2D spectral coefficients therefore retain the three-dimensional field information required by the particle pusher. Figure \ref{cgtb_6n} shows the timing breakdown for multi-$n$ case with 2000 time steps on the same laptop used for the single-$n$ benchmark. The 1-GPU/1-MPI simulation takes 272.9~s (4.5~min), whereas the 4- and 8-rank CPU simulations take 1395.7 and 1069.1~s, respectively. The six toroidal mode linear simulation can complete in less than 5~min on the laptop GPU and in approximately 18--23~min on the laptop CPU.

\begin{figure}[H]
	\center
	\includegraphics[width=1.0\textwidth]{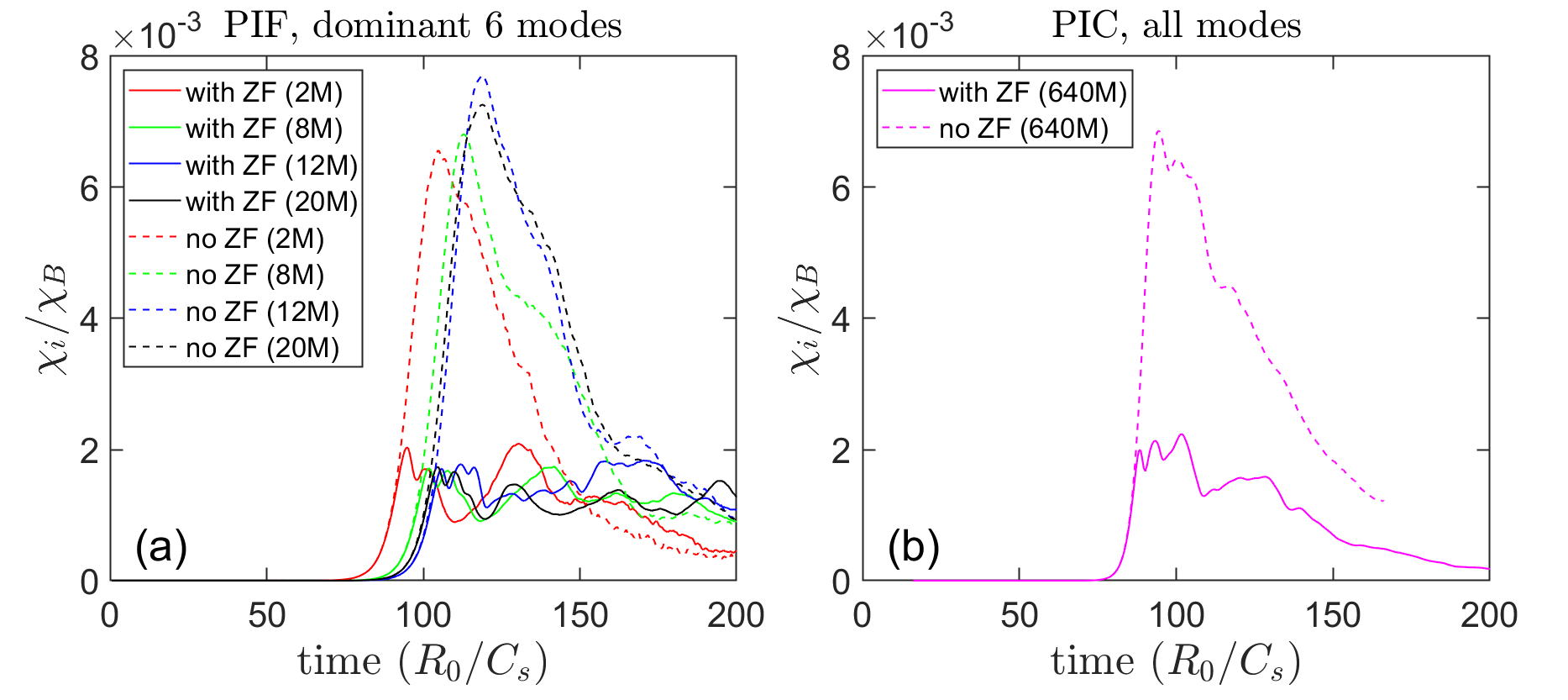}
	\caption{Time histories of the ion heat conductivity $\chi_i/\chi_B$ with and without the zonal-flow (ZF) component. (a) Hybrid spectral PIF simulations retaining the six dominant toroidal modes $n=13$--18, with total marker populations of 2, 8, 12, and 20 million. (b) Conventional PIC simulations retaining all toroidal modes, with approximately 640 million markers. Here, $\chi_B=cT_e/(eB_a)$ is the Bohm diffusivity.}
	\label{nIh}	
\end{figure}

For the nonlinear ITG turbulence calculation, the hybrid spectral PIF method retains the six dominant toroidal modes $n=13$--18. Figure \ref{nIh}(a) compares the normalized ion thermal diffusivity $\chi_i/\chi_B$ for cases with and without the zonal-flow component and with total marker populations of 2, 8, 12, and 20 million. The cases have similar linear growth and show the same reduction of the nonlinear transport level by zonal flow. In particular, the 2 million marker result is very close to the 20 million marker result, indicating that the main nonlinear transport behavior is already resolved with 2 million markers. The conventional PIC calculation in figure \ref{nIh}(b), for comparison, retains all toroidal modes and uses approximately 640 million markers to obtain a similar transport level. The time history comparison of ITG transport coefficient between cases with and without zonal flow is consistent with the large-scale GTC results in Ref. \cite{Lin98}.

\begin{figure}[H]
	\center
	\includegraphics[width=1.0\textwidth]{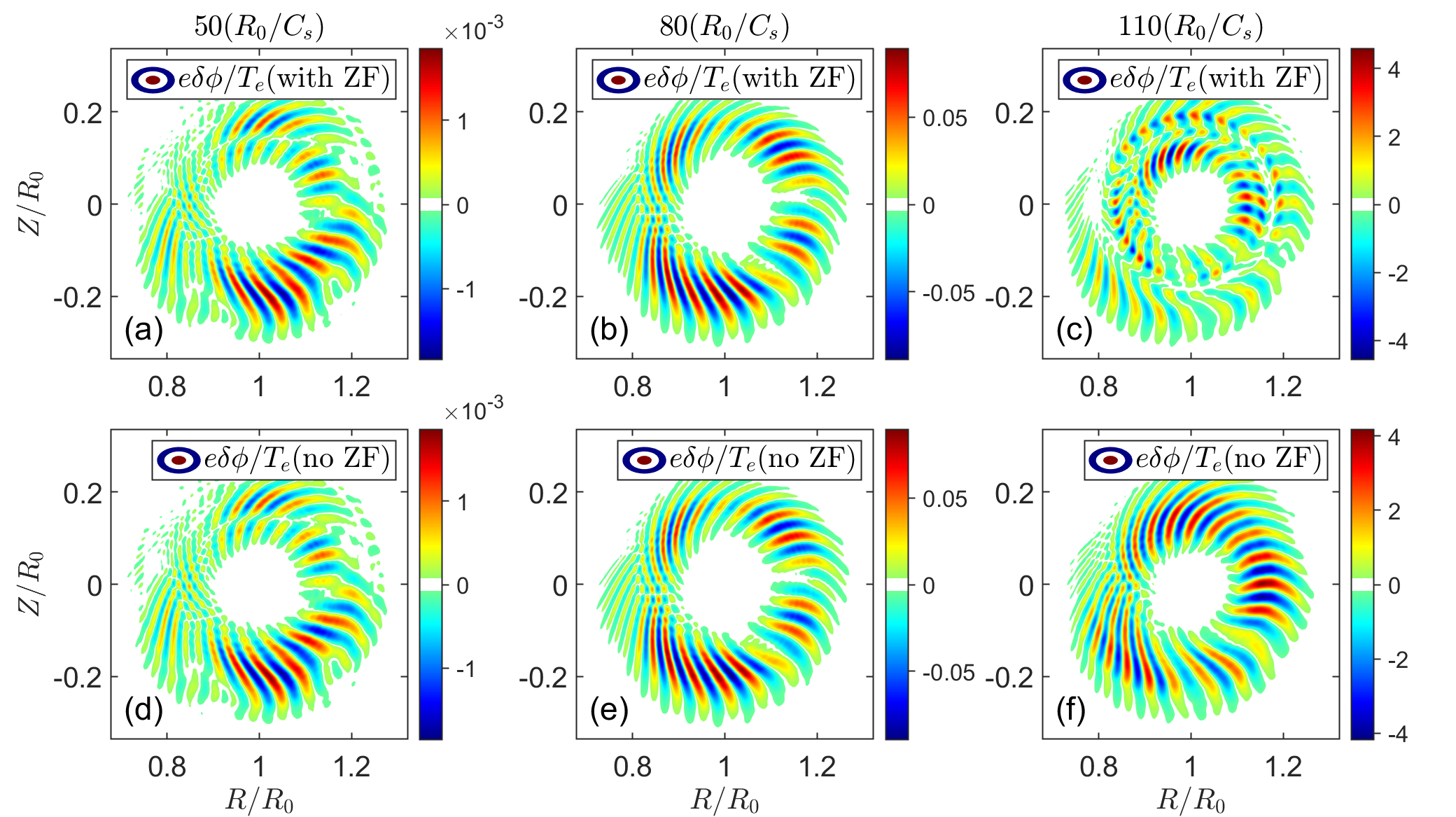}
	\caption{Snapshots of the normalized electrostatic-potential perturbation $e\delta\phi/T_e$ from the 20 million marker hybrid spectral PIF simulations with (upper row) and without (lower row) zonal flow at $t=50R_0/C_s$, $80R_0/C_s$, and $110R_0/C_s$ in figure \ref{nIh}, from left to right columns.}
	\label{nIs}	
\end{figure}

The $\delta\phi$ snapshots in figure \ref{nIs} illustrate the regulation process of ITG turbulence by the self-generated zonal flow. At $t=50R_0/C_s$ and $80R_0/C_s$, the cases with and without zonal flow have similar ballooning structures. Around $t=110R_0/C_s$, the case with zonal flow has obvious radial distortion and much weaker radial coherence, whereas radially extended turbulence eddies remain in the case without zonal flow. This difference is consistent with turbulence regulation by zonal flow $E\times B$ shear \cite{Lin98}. For the 2 million marker run with zonal flow, 4000 time steps covering $200R_0/C_s$ take 572.2~s on one laptop GPU with one MPI rank. According to figure \ref{nIh}, this result shows that the long time nonlinear dynamics can also be obtained within minutes.

\subsection{GPU scaling performance}\label{scaling}

We next examine preliminary multi-GPU performance on NVIDIA A100 GPUs. Conventional toroidally decomposed PIC simulations require particle-shift communication when markers cross subdomain boundaries \cite{Zhang2018,Either2005}, this communication is not required from the hybrid spectral PIF method. In the scaling test, we vary the GPU number $N_g$, marker number $N_p$, and retained toroidal-mode number $N_n$. Each case is advanced for 2000 time steps with one MPI rank per GPU.

\begin{figure}[H]
	\center
	\includegraphics[width=0.9\textwidth]{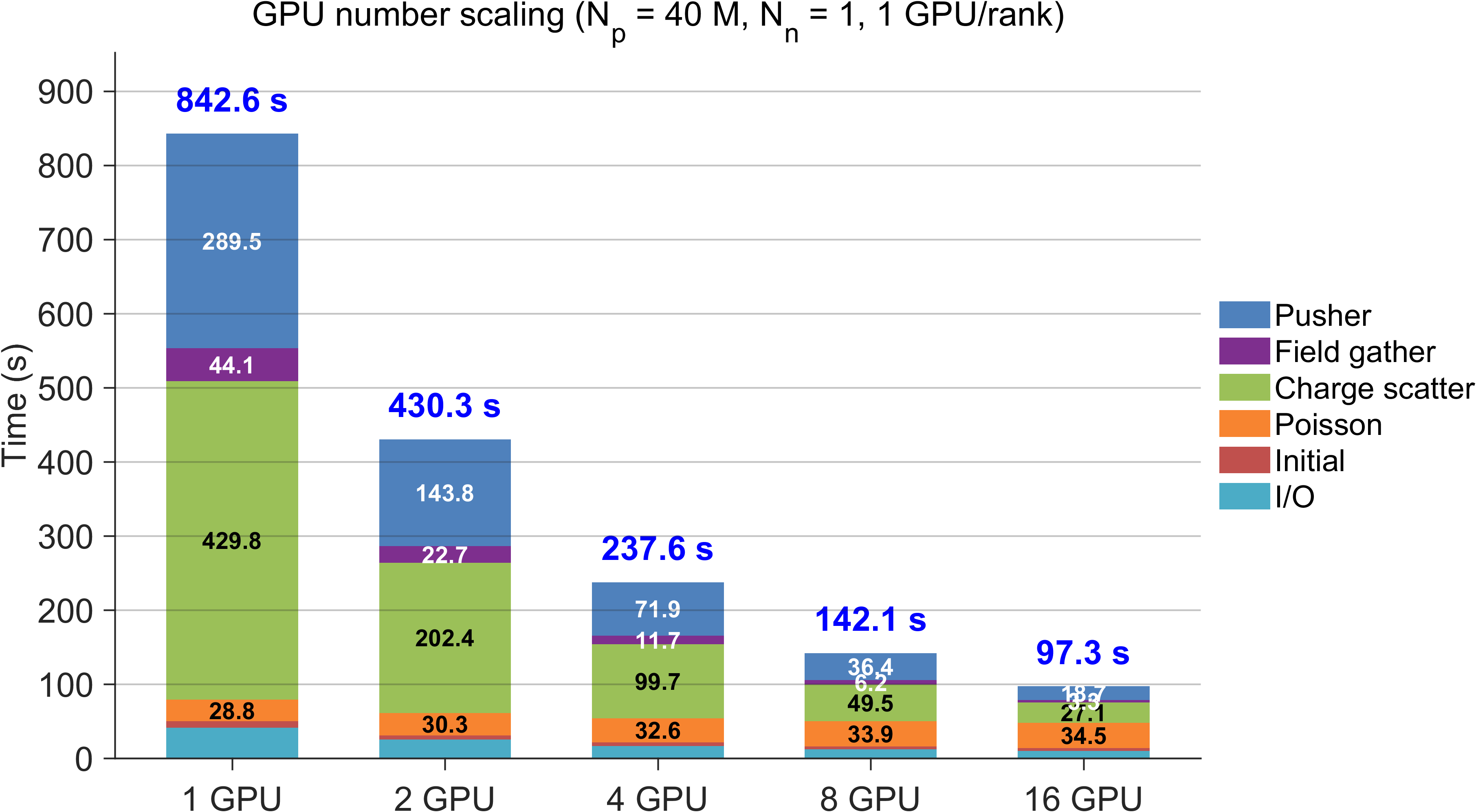}
	\caption{Strong scaling runtime breakdown for $N_p=40$ million and $N_n=1$ on 1, 2, 4, 8, and 16 NVIDIA A100 GPUs (2000 time steps), with one MPI rank per GPU. The total runtime is shown above each bar, and the principal runtime components are labeled within the bars.}
	\label{tsn}	
\end{figure}

\begin{figure}[H]
	\center
	\includegraphics[width=0.9\textwidth]{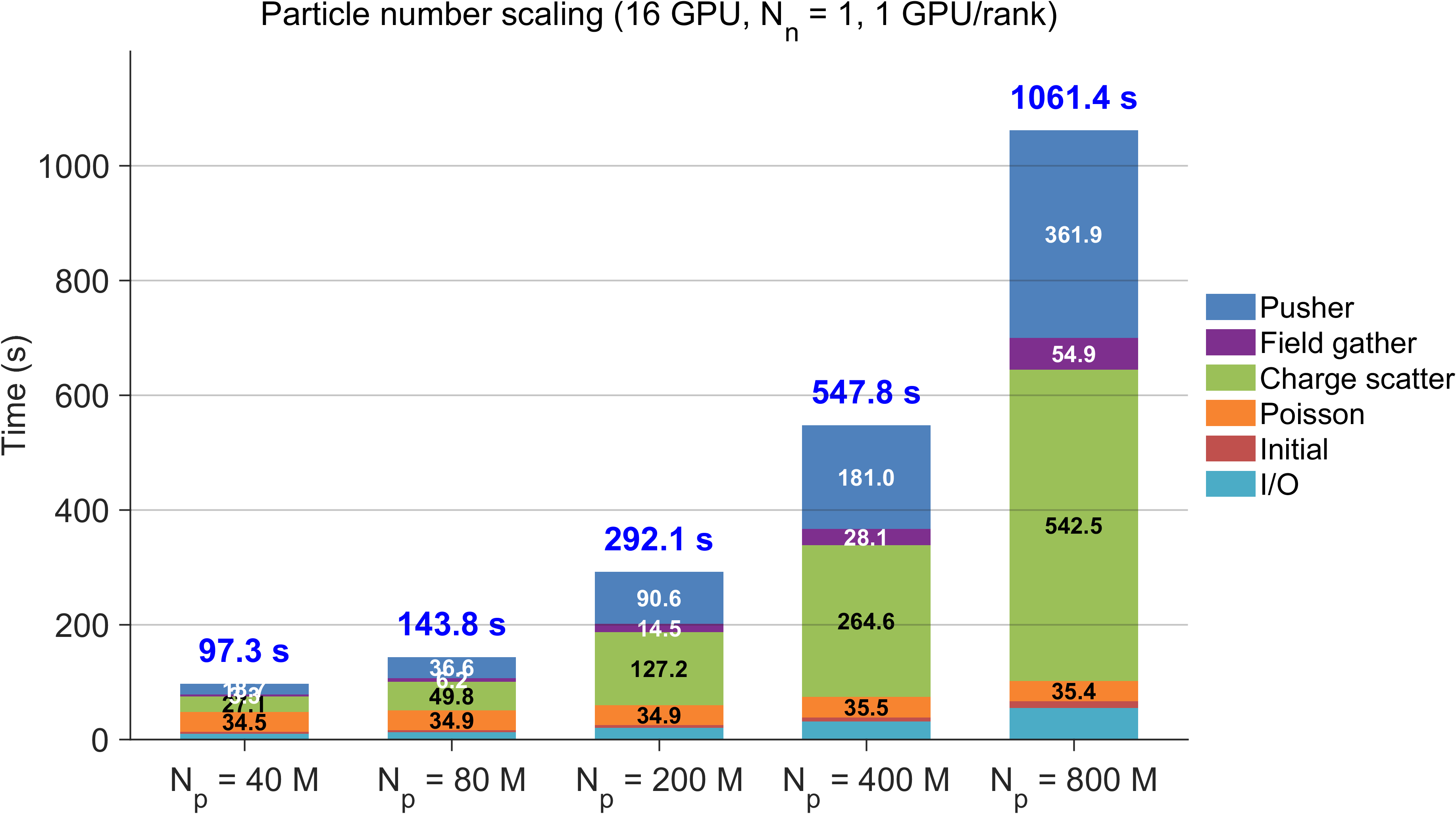}
	\caption{Runtime breakdown at fixed $N_g=16$ and $N_n=1$. $N_p=40$, 80, 200, 400, and 800 million. Each case is advanced for 2000 time steps on NVIDIA A100 GPUs with one MPI rank per GPU. The total runtime is shown above each bar, and the principal runtime components are labeled within the stacked bars.}
	\label{tsp}	
\end{figure}

\begin{figure}[H]
	\center
	\includegraphics[width=0.9\textwidth]{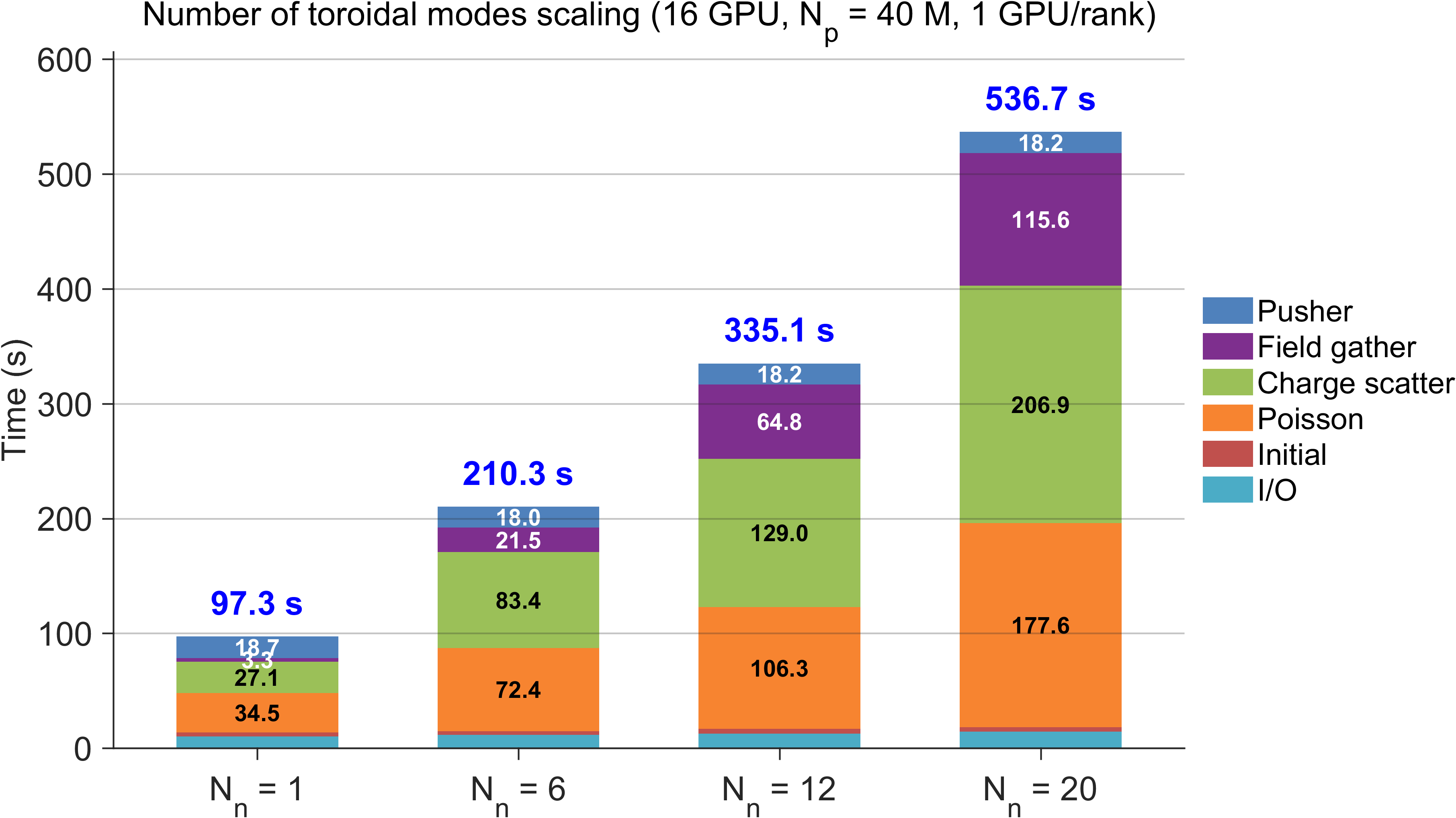}
	\caption{$N_n$ scaling at fixed $N_g=16$ and $N_p=40$ million, $N_n=1$, 6, 12, and 20. Each case is advanced for 2000 time steps on NVIDIA A100 GPUs with one MPI rank per GPU. The total runtime is shown above each bar, and the principal runtime components are labeled within the stacked bars.}
	\label{tsnn}	
\end{figure}

Figure \ref{tsn} shows a strong-scaling test with fixed $N_p=40$ million and $N_n=1$. Increasing $N_g$ from 1 to 16 reduces the total runtime from 842.6 to 97.3~s, a speedup factor of 8.66 on-average, this is due to the fact that Poisson solver is solved using 1 MPI in this single-n test, and could be further accelerated with more MPI ranks. The particle pusher, field gather, and charge scatter times decrease nearly in proportion to $1/N_g$, resulting a strong-scaling parallel efficiency of 100\% for these individual key parts of particle simulation, which also indicates the efficient scaling on multiple GPUs without the delay of toroidal particle-shift communication \cite{Either2005}. The Poisson-solver time, however, remains at approximately 30 s because the reduced matrix has the same size as the single-$n$ case and is solved on one GPU. Figure \ref{tsp} shows a marker number scan at $N_g=16$ and $N_n=1$, rather than a conventional weak-scaling test. As $N_p$ increases from 40 to 800 million, the particle pusher and charge scatter times increase almost in proportion to $N_p$, and field gather follows the same trend. The Poisson-solver time changes little because the field problem is independent from $N_p$. Increasing the marker number introduces no additional toroidal particle-shift communication cost. Figure \ref{tsnn} shows the number of toroidal mode scan at fixed $N_g=16$ and $N_p=40$ million. The particle pusher time remains nearly unchanged as $N_n$ increases from 1 to 20 as expected. Field gather and charge scatter times increase with $N_n$, since each toroidal mode need to be calculated for all markers, which are not parallelized for different $n$ in this paper. The Poisson time increases much slower than $N_n$ which distributes different $n$ components among MPI ranks. In a separate work, the multi-\(n\) optimization is performed by fully distributing independent $n$-mode systems across MPI ranks in batches, with each rank solving one \(n\) at a time, which lower memory and communication overhead and improve strong scaling. MPI shared-memory allow ranks on the same node to reuse selected particle, grid, and field information instead of storing duplicate copies. The charge-scatter and field-gather stages use reduced scatter/gather communication and exchange only the data required by each mode. Allreduce operations are limited to essential global quantities and synchronization, reducing both the number of use and data size.

\section{Conclusion}

A hybrid spectral method is formulated within the particle-in-Fourier (PIF) framework and implemented in the electrostatic model of GTC. By converting the conventional three-dimensional particle-grid problem to lower-dimensional spectral representations, the method reduces the effective problem size by a factor of 48 and achieves a speedup of over two orders of magnitude for single-$n$ simulation. As a result, high-fidelity electrostatic gyrokinetic simulations can be completed on a laptop within minutes. Preliminary scaling tests also demonstrate effective performance across multiple NVIDIA A100 GPUs on supercomputer.

The method employs different spectral representations for charge scatter, field gather, and the Poisson solver. Particle-grid coupling is performed using toroidal Fourier coefficients on a 2D poloidal mesh, while the Poisson solver retains only the physically relevant coupled $m$-harmonics in a reduced sparse matrix. Truncated spectral transforms connect these representations and avoid too many particle-grid operations for each individual $m$-harmonic, thereby significantly reducing the cost of particle-field interpolation processes. Single-$n$ benchmarks show excellent agreement with conventional PIC on both mode structures and dispersion relations. Multi-$n$ simulations also recover nonlinear ion heat transport and zonal flow regulation, consistent with well-established large-scale GTC results. A 2000-step single-$n$ simulation with approximately 2 million markers completes in 78.2 seconds on a laptop GPU, and multi-$n$ turbulence simulations can finish within minutes. In scaling tests, the 40-million-marker case achieves a strong-scaling parallel efficiency of 100\% for several key parts of particle simulation, including particle pusher, charge scatter and field gather. 

We next plan to extend the hybrid spectral PIF method to electromagnetic simulations, and develop a high-fidelity gyrokinetic solver that can effectively mitigate the well-known "cancellation problem" associated with kinetic electrons \cite{Bao2017, Bao2018, Lu2025}. Future upgrades will also incorporate finite-element upgrades \cite{Lu2026a, Lu2026b} and address the singularity near the magnetic axis \cite{Jiang2026}. Together with the hybrid spectral PIF method presented in this work, these extensions are essential for enabling cross-scale simulations that encompass microturbulence, energetic-particle-driven instabilities, and macroscopic MHD modes.

\section{Acknowledgments}
The authors would like to thank Prof. Lai Wei, Prof. Zhengxiong Wang, Chen Zhao, Peiyou Jiang, Youjun Hu, and Yueyan Li for helpful discussions.

	\end{sloppypar}
\end{document}